\documentstyle[12pt,rotate]{article}

\input epsf.sty

\newcommand{\ft}[2]{{\textstyle\frac{#1}{#2}}}

\newcommand{\re}{{\rm e}}

\newcommand{\OO}{\mathop{\otimes}}
\font\cmss=cmss12 
\def\1{\hbox{{1}\kern-.25em\hbox{l}}}
\def\bfZ{\relax{\hbox{\cmss Z\kern-.4em Z}}}
\newcommand{\qqv}{{^{QQ}\! f^T}}
\newcommand{\qqva}{{^{QQ}\! f^a}}
\newcommand{\qqbv}{{^{QQ} \bar f^T}}
\newcommand{\HQQ}{{^{QQ}\! h}}
\newcommand{\bHQQ}{{^{QQ} \bar h}}
\newcommand{\ggvv}{{^{GG}\! f^T}}
\newcommand{\ggbvv}{{^{GG} \bar f^T}}
\newcommand{\ggvc}{{^{GG}\! f^c}}

\newcommand{\ggva}{{^{GG}\! f^a}}
\newcommand{\HGGe}{{^{GG}\! h^T}}
\newcommand{\bHGGe}{{^{GG} \bar h^T}}

\newcommand{\beq}{\begin{equation}}
\newcommand{\eeq}{\end{equation}}
\newcommand{\bea}{\begin{eqnarray}}
\newcommand{\eea}{\end{eqnarray}}

\newcommand{\bz}{\bar z}
\newcommand{\be}{\bar \zeta}
\newcommand{\zoe}{\frac{z}{\zeta}}
\newcommand{\bzoe}{\frac{\bar z}{\bar \zeta}}

\newcommand{\embze}{\left ( 1 - \frac{\bar z}{\bar \zeta}\right )}
\newcommand{\lozembze}{\ln^2 \left ( 1 - \frac{\bar z}{\bar \zeta}\right )}
\newcommand{\loembze}{\ln \left ( 1 - \frac{\bar z}{\bar \zeta}\right )}

\newcommand{\loe}{\ln \bar \zeta}
\newcommand{\lozz}{\ln^2 z}
\newcommand{\loz}{\ln z}
\newcommand{\lobz}{\ln \bar z}

\newcommand{\lobzobe}{\ln\left(\frac{\bar z}{\bar \zeta}\right)}

\newcommand{\pqqt}{{^{QQ}p^T}}
\newcommand{\pqqa}{{^{QQ}p^a}}
\newcommand{\pggpt}{{^{GG}\!p^T}}
\newcommand{\pgga}{{^{GG}\!p^a}}
\newcommand{\pggc}{{^{GG}\!p^c}}

\begin{document}
\begin{titlepage}

\centerline{\large \bf NLO evolution kernels for skewed transversity
                    distributions.}

\vspace{15mm}

\centerline{\bf A.V. Belitsky$^{a,c}$, A. Freund$^b$, D. M\"uller$^c$}

\vspace{10mm}

\centerline{\it $^a$C.N.\ Yang Institute for Theoretical Physics}
\centerline{\it State University of New York at Stony Brook}
\centerline{\it NY 11794-3840, Stony Brook, USA}

\vspace{5mm}

\centerline{\it $^b$INFN, Sezione di Firenze, Largo E. Fermi 2}
\centerline{\it 50125 Firenze, Italy}

\vspace{5mm}

\centerline{\it $^c$Institute f\"ur Theoretische Physik,
                Universit\"at Regensburg}
\centerline{\it D-93040 Regensburg, Germany}

\vspace{15mm}

\centerline{\bf Abstract}

\vspace{0.3cm}

\noindent
We present a calculation of the two-loop evolution kernels of the twist-two
transversally polarized skewed quark and linearly polarized skewed gluon
distributions in the minimal subtraction scheme and discuss a solution of
the evolution equations suitable for numerical implementation.

\vspace{6cm}


\vspace{0.5cm}


\end{titlepage}

%%%%%%%%%%%%%%%%%%%%%%%%%%%%%%%%%%%%%%%%%%%%%%%%%%%%%%%%%%%%%%%%%%%%%
\section{Introduction.}
\label{sec-int}
%%%%%%%%%%%%%%%%%%%%%%%%%%%%%%%%%%%%%%%%%%%%%%%%%%%%%%%%%%%%%%%%%%%%%

Helicity-flip parton densities, i.e.\ densities of transversally polarized
quarks and linearly polarized gluons, are inaccessible in conventional deep
inelastic scattering experiments on polarized spin-$\ft12$ targets. The
transversally polarized quark density \cite{RalSop79} is of odd chirality,
and thus requires a helicity flip in a perturbative QCD subprocess
which is forbidden for massless quarks, while the linearly polarized gluon
requires the helicity to be flipped by two units which is not allowed by
angular momentum conservation for spin-$\ft12$ hadrons. However, it will
definitely appear for spin-$J \geq 1$ hadrons/photon
\cite{JafMan89,ErmKirSzy99}.

If one goes from the forward to off-forward kinematics, i.e.\
inclusive DIS to exclusive DVCS with a real final state photon
\cite{MueRobGeyDitHor94DitGeyMueRobHor88,Ji97,Rad97}, one allows for non-zero
orbital momentum in the system and thus the gluons can transfer two units
of helicity from photons to spin-$\ft12$ hadrons
\cite{Diehl97,HooJi98,BelMue00}. Due to specific $\cos/\sin
\left( 3 \phi \right)$ azimuthal angle dependences in the
factorized \cite{Rad97,ColFre99,JiOsb98} cross section of the
corresponding skewed tensor gluon distribution (SPD), it can be
extracted \cite{BelMue00} without the usual complication due to quark
contamination. A similar possibility exists for a system of hadrons due
to their orbital motion, as it occurs in the $\gamma\gamma \to \pi\pi$
reaction \cite{KivManPol99}. As to the quark sector, unfortunately,
the original proposal to measure the skewed quark transversity
in diffractive meson production \cite{ColFraStr97} has been proven erroneous
due to the preservation of chiral symmetry for the cross section
\cite{ColDie00} and thus complicates the experimental access to
the function in question.

Note, however, in DVCS, since the photon helicity flip requires a
change in the hadron helicity by two units, the amplitude on a proton target
is suppressed (due to its spin being only $\ft12$) by an extra power of the 
transverse momentum transfer in the $t$-channel $\Delta_\perp$ whereas it will
not be suppressed on a deuteron target, which is available at certain facilities
like HERMES. Given its current run on high density targets, it is very interesting that high
statistics data will be shortly forthcoming and hence there is a need to
study the helicity-flip amplitudes theoretically.

In this paper, we present an evaluation of the two-loop skewed evolution kernels
for chiral odd quark and tensor gluon distributions. Since
corresponding operators which define the non-perturbative content of the
former belong to different representations of the Lorentz group the mixing
is absent in both cases. Our presentations is structured as follows. In the
second section, we discuss the general solution of the skewed evolution
equations, in the third one, we present the reconstruction of the skewed
evolution kernels in the Efremov-Radyushkin-Brodsky-Lepage (ER-BL) and
Dokshitzer-Gribov-Lipatov-Altarelli-Parisi (DGLAP) regions, and finally, we
summarize.

%%%%%%%%%%%%%%%%%%%%%%%%%%%%%%%%%%%%%%%%%%%%%%%%%%%%%%%%%%%%%%%%%%%%%
\section{Evolution equation.}
\label{sec-evo}
%%%%%%%%%%%%%%%%%%%%%%%%%%%%%%%%%%%%%%%%%%%%%%%%%%%%%%%%%%%%%%%%%%%%%

SPDs \cite{MueRobGeyDitHor94DitGeyMueRobHor88}-\cite{Rad97,ColFraStr97}
appear in
various exclusive processes and their extraction would provide us new,
valuable, non-perturbative information on the structure of hadrons. These
functions can be considered as a link between Feynman's parton densities and
exclusive quantities like form factors/distribution amplitudes, in fact they
are generalizations of both. In QCD they are defined as expectation values
of twist-two light-ray operators over the hadronic states,
\begin{eqnarray}
\label{def-SPD-tra}
\left\{{ {^{Q}\! q^{T}} \atop {^{G}\! q^{T}} }
\right\}(x, \eta, \Delta^2; Q^2)
=
\left\{{ 1 \atop 4 P_+^{-1} } \right\}
\int \frac{d \kappa}{2 \pi} {\rm e}^{i \kappa x P_+}
\langle P_2 |
\left\{ {^Q\! {\cal O}^T \atop {^G\! {\cal O}^{T}} } \right\}
(\kappa, -\kappa)_{| \mu^2 = Q^2}
| P_1 \rangle,
\end{eqnarray}
where $P_+ = n \cdot (P_1 + P_2)$ with $n_\mu$ being a light cone vector.
These are functions of the momentum fraction $x$, the longitudinal
momentum fraction $\eta > 0$ in the $t$ channel, called the skewedness
parameter, the momentum transfer square $\Delta^2 = (P_2 - P_1)^2$,
and they vary depending on the resolution scale $Q^2$. The partonic
interpretation of these functions depends on the kinematical region.
For the ER-BL region $|x| < \eta$ they behave as distribution
amplitudes of meson-like states, while for the DGLAP domain $|x| > \eta$
they mimic the conventional probability distribution of finding a
parton with a corresponding momentum and spin state that is
scattered from the initial to the final state.

Now we recall the general properties of the evolution equation for SPDs,
\begin{eqnarray}
\label{Def-EvoEquSPD}
\frac{d}{d \ln Q^2} \, {^A\! q^T} (x, \eta, \Delta^2; Q^2)
= \int_{-1}^{1} \frac{dy}{2|\eta|} \,
{^{AA}V^T}
\left(
\frac{\eta + x}{2\eta}, \frac{\eta + y}{2\eta}; \alpha_s (Q^2)
\right)
{^A\! q^T} (y, \eta, \Delta^2; Q^2) ,
\end{eqnarray}
with the kernel ${^{AA}V^T}$ defined as a series in the coupling
$\alpha_s$. Note, that we treat both quark and gluon cases simultaneously and that 
there is no mixing between them, as
mentioned in the introduction, since they belong to different
representations of the Lorentz group.

The evolution kernel ${^{AA}V^T} (x, y)$ with $0 \le x, y \le 1$ is of
the ER-BL-type and governs the evolution of meson distribution amplitudes
(formally $|\eta| \equiv 1$). The general structure of this kernel was
already studied in the past \cite{MueRobGeyDitHor94DitGeyMueRobHor88}
with the main result being that the evolution kernels of skewed parton
distributions can be obtained from the ER-BL ones,
\begin{eqnarray}
\label{def-ERBL}
{^{AA}V^T} = \theta(y - x) f(x, y) + \theta( y - \bar x) g (x, y)
+ \left\{{ x \to \bar x \atop y \to \bar y} \right\},
\qquad \bar x \equiv 1 - x,
\qquad 0 \le x, y \le 1 ,
\end{eqnarray}
[here $x$ denotes a new variable, which differs from that one used in
Eq.\ (\ref{Def-EvoEquSPD})] by replacing the $\theta$ structure:
\begin{eqnarray}
\label{Extension}
\theta (y - x)
\to
\Theta \left( \frac{\eta + x}{2\eta}, \frac{\eta + y}{2 \eta} \right)
\equiv
\theta \left( 1 - \frac{\eta + x}{\eta + y} \right)
\theta \left( \frac{\eta + x}{\eta + y} \right)
\mbox{sign} \left( \frac{\eta + y}{\eta} \right),
\end{eqnarray}
for $x \to \ft1{2\eta}( \eta + x )$, and analytical continuation of the
prefactors $f$ and $g$. In the gluonic case the second $\theta$ structure
can be reduced to the first one by means of Bose symmetry, namely,
${^G\! q^T} (x, \eta) = {^G\!q^T} (- x, \eta)$, while in the quark case,
we should decompose the SPD in contributions\footnote{More precisely, we
should speak of (anti) quark distributions only for $x \ge \eta$ and
($x \le - \eta$), while there are two meson-like distributions with
different symmetry in the ER-BL region.} coming from quarks ($x \ge 0$)
and anti-quarks ($x \le 0$).

The evolution equation (\ref{Def-EvoEquSPD}) can be solved by means
of different methods. However, most of them are plagued by either a
limitation of their applicability to a leading order (LO) analysis or a lack
of sufficient accuracy in handling particular regions of the
phase space. Only the orthogonal polynomial reconstruction method
\cite{BelGeyMueSch97,ManPilWei97a} and the direct numerical integration
methods \cite{FraFreGuzStr97,MusRad99} allow for a generalization beyond the
one-loop approximation. However, the first method suffers from
numerical complications in the treatment of the transition region $|x| = \eta$
and, therefore, one may explore the efficiency of a direct numerical
integration to overcome this particular difficulty. To do so we should separate the
evolution equation into DGLAP and ER-BL regions.

In the quark sector, it proves convenient to introduce (non-) singlet
like combinations of SPDs that are symmetric/antisymmetric w.r.t.\
the momentum fraction $x$:
\begin{eqnarray}
{^Q\!q^T} (x, \eta)
= \ft12 \left[ {^+\!q^T} (x, \eta) - {^-\!q^T} (- x, \eta) \right] ,
\qquad
{^\pm\!q^T} (x, \eta) = {^Q\!q^T} (x, \eta) \mp {^Q\!q^T} (- x, \eta) .
\end{eqnarray}
Here the antiquark contribution is given by $-{^Q\!q^T} (- x, \eta)$
and thus ${^{\pm}\!q^T} (x, \eta)$ for $x > 0$ is just the sum and
difference of quark and anti-quark distributions, respectively. The
kernels in the corresponding evolution equations read
\begin{eqnarray}
{^{QQ}V^{T \pm}}
\left( \frac{\eta + x}{2\eta}, \frac{\eta + y}{2 \eta} \right)
\!\!\!&=&\!\!\!
\Theta \left( \frac{\eta + x}{2\eta}, \frac{\eta + y}{2\eta} \right)
{^{QQ}\!F^{T \pm}}
\left( \frac{\eta + x}{2\eta}, \frac{\eta + y}{2\eta} \right)
+ \left\{ { x \to - x \atop y \to - y } \right\} , \\
{^{QQ}\!F^{T \pm}} (x, y)
\!\!\!&=&\!\!\! {^{QQ}\!f^T} (x, y) \mp {^{QQ}\!g^T} ( \bar x, y) .
\nonumber
\end{eqnarray}

Based on the support and symmetry properties, we discuss now the
solution of this evolution equation (\ref{Def-EvoEquSPD}). In the
following ${^\pm\!q} (x, \eta, Q^2)$ is a (anti-) symmetric skewed
distribution of any species, where for gluons we formally have to
take the solution for ${^-\!q} (x, \eta, Q^2)$. An analogous sign
convention holds true for unpolarized partons, while for longitudinal
polarized ones, one should replace $+ \leftrightarrow -$. In the
DGLAP region we have a homogeneous integro-differential equation,
\begin{eqnarray}
\label{Def-EvoEqaDGLAP}
\frac{d}{d \ln Q^2} {^{\pm}\!q} (x, \eta; Q^2)_{|x| > \eta}
= \int_{x}^{1} \frac{dy}{2|\eta|}
\, \left( {^{\pm}\!F} - {^{\pm}\!\overline{F}} \right)
\left( \frac{\eta + x}{2\eta}, \frac{\eta + y}{2\eta} \right)
{^{\pm}\!q} (y, \eta; Q^2) ,
\end{eqnarray}
while in the ER-BL region an inhomogeneous term arises,
\begin{eqnarray}
\label{Def-EvoEqaERBL}
\frac{d}{d \ln Q^2} {^{\pm}\!q} (x, \eta; Q^2)_{|x| < \eta}
\!\!\!&=&\!\!\! \int_{-|\eta|}^{|\eta|} \frac{dy}{2|\eta|} \,
{^{\pm}V}
\left( \frac{\eta + x}{2\eta}, \frac{\eta + y}{2\eta} \right)
{^{\pm}\!q} (y, \eta; Q^2) + {^{\pm}\!I} (x, \eta; Q^2) , \\
{^{\pm}\!I} (x, \eta; Q^2)
\!\!\!&=&\!\!\! \int_{|\eta|}^1 \frac{dy}{2|\eta|} \,
\left[
{^{\pm}\!F} \left( \frac{\eta + x}{2\eta}, \frac{\eta + y}{2\eta} \right)
\mp
{^{\pm}\!F} \left( \frac{\eta - x}{2\eta}, \frac{\eta + y}{2\eta} \right)
\right]
{^{\pm}\!q} (y, \eta; Q^2) ,
\nonumber
\end{eqnarray}
where we used the short-hand notation $\overline{F}(x,y) \equiv F ( \bar x,
\bar y )$. Introducing the kernel $P \left( \ft{x}{y}, \ft{\eta}{y} \right)
= \ft{y}{|\eta|} \left( F - \overline{F} \right) \left( \frac{\eta +
x}{2\eta}, \frac{\eta + y}{2\eta} \right)$, it is obvious that equation
(\ref{Def-EvoEqaDGLAP}) is of the DGLAP type. Since, in addition, the kernel depends
on the skewedness parameter $\eta$, there does not exist the same
factorization for the Mellin moments as in the forward case. Fortunately,
this equation can be efficiently integrated by brute force
\cite{FraFreGuzStr97}. We formally write the solution of this equation as
a convolution of the skewed distribution given at the input scale $Q_0^2$ with
an evolution operator
\begin{eqnarray}
\label{Sol-DGLAP}
q (x, \eta; Q^2)_{|x| > \eta} = \int_{|\eta|}^{1} \frac{dy}{y} \,
U_{\rm DGLAP} \left( x, y, \eta; Q^2, Q_0^2 \right) q (y, \eta; Q_0^2) ,
\end{eqnarray}
that satisfies the equation
\begin{eqnarray}
\frac{d}{d \ln Q^2} U_{\rm DGLAP} (x, z, \eta; Q^2, Q_0^2)
&=& \int_{x}^1 \frac{dy}{y} P \left( \frac{x}{y}, \frac{\eta}{y} \right)
U_{\rm DGLAP} (y, z, \eta; Q^2, Q_0^2)
\end{eqnarray}
with the initial condition $U_{\rm DGLAP} \left(x, y, \eta; Q_0^2, Q_0^2
\right) = \delta \left( 1 -\ft{x}{y} \right)$. The homogeneous part of
the evolution equation (\ref{Def-EvoEqaERBL}) is precisely of the ER-BL
type and its solution is
\begin{equation}
q^{\rm hom} (x, \eta; Q^2)_{|x| < \eta} = \int_{-1}^{1} dy \,
U_{\rm ERBL} \left( \frac{x}{\eta}, y; Q^2, Q_0^2 \right)
q^{\rm hom} (\eta y, \eta; Q_0^2) ,
\end{equation}
with $U_{\rm ERBL} \left(x, y, Q_0^2, Q_0^2 \right) = \delta(x - y)$.
A particular solution for the inhomogeneous equation is easily constructed
by means of the inverse evolution operator and we obtain, together with
the homogeneous part,
\begin{eqnarray}
q (x, \eta; Q^2)_{|x| < \eta}
\!\!\!&=&\!\!\! \int_{-1}^{1} dy \, U_{\rm ERBL}
\left( \frac{x}{\eta}, y; Q^2, Q_0^2 \right)
\Bigg\{ q( \eta y, \eta; Q_0^2 ) \nonumber\\
&+&\!\!\! \int_{-1}^1 dz \, \int_{Q_0^2}^{Q^2} \frac{dQ'^2}{Q'^2} \,
U^{-1}_{\rm ERBL} \left( y, z; Q'^2, Q_0^2 \right)
I (\eta z, \eta; Q'^2) \Bigg\} ,
\end{eqnarray}
where $I (x, \eta; Q'^2) $ is the convolution of Eq. (\ref{Def-EvoEqaERBL})
with the solution of Eq.\ (\ref{Sol-DGLAP}) in the DGLAP region. An
important issue is to study the evolution at the point $|x| = \eta$. Since
we have the property $F (x = 0, y) = \overline{F} (x = 1, y) = 0$ in both LO
and next-to-leading (NLO) [see below Eqs.\ (\ref{Def-ERBL-LO-tr}),
(\ref{Def-ERBL-NLO-trQ}), and (\ref{Def-ERBL-NLO-trG})], the limit $\epsilon
\to 0$ of the evolution equations for $x = \eta + \epsilon$ and $x = \eta -
\epsilon$ will be the same. Therefore, one expects that a smooth input
function will remain smooth under evolution. Nevertheless, this point necessitates
care as it is a delicate point in the numerical solution of the
evolution equations.

The construction of the ER-BL from the DGLAP kernels is in principle
possible, provided we know the eigenfunctions of the former. In LO the
exclusive kernels are diagonal w.r.t.\ Gegenbauer polynomials
\begin{eqnarray}
\label{eigenvec}
\int_{-1}^1 \frac{dx}{2|\eta|} \,
C_{j + 3/2 - \nu(A)}^{\nu(A)} \left( \frac{x}{\eta} \right)
{^{AA}V^{(0)T}}
\left( \frac{\eta + x}{2\eta}, \frac{\eta + y}{2\eta} \right)
= - \frac{1}{2} {^{AA}\gamma_j^{(0)T}}
C_{j + 3/2 - \nu(A)}^{\nu(A)} \left( \frac{y}{\eta} \right) ,
\end{eqnarray}
where $\nu(A) = \left\{\ft32, \ft52 \right\}$ for $A = \{ Q, G \}$ and
${^{AA}\!\gamma_j^{(0)T}}$ are the forward anomalous dimensions at LO.
This fact reflects the underlying conformal symmetry at tree-level.
Consequently, this symmetry allows us to reconstruct the ER-BL kernel
from the DGLAP kernel by an integral transformation [for details see
second article of \cite{Mue94BelMue98aBelMue98c}],
${^{AA}V^{(0)T}}(x,y) = \int_0^1 dz\; G(2x-1, 2y-1; z| \nu(A))
{^{AA}P^{(0)T}}(z)$,
where the integral kernel is given in terms of a hypergeometric function
\begin{eqnarray}
\label{Def-IntKerG}
G(x, y; z| \nu) 
&=&
\frac{\Gamma(\nu)\Gamma(\nu + 1)}{\Gamma^2(\frac{1}{2})\Gamma(2\nu)}
\frac{2^{2\nu}(1-x^2)^{\nu -1/2}(1 - z^2)}{
\left[ 1 - 2 \left(x y -\sqrt{(1-x^2)(1-y^2)} \right) z + z^2
\right]^{\nu + 1}}
\\
&&\times
{_2F_1}
\left( { \nu + 1, \nu \atop 2 \nu }
\left| \frac{4\sqrt{(1-x^2)(1-y^2)} z}{1
- 2 \left( x y - \sqrt{(1-x^2)(1-y^2)}\right) z + z^2 } \right. \right).
\nonumber
\end{eqnarray}
Note that, in any order of perturbation theory, the evolution operator
in the ER-BL region can be represented in terms of an infinite series of
Gegenbauer polynomials \cite{Mue95}. Unfortunately, since in general a
SPD does not vanish at the point $x = \pm \eta$, it is expected that, to
achieve a good approximation, a large number of terms in this expansion
is needed. Fortunately, we can resum the Gegenbauer expansion in the ER-BL
region which provides us the integral transformation, mentioned above,
of the evolution operator in the forward region
\begin{eqnarray}
U_{\rm ERBL}\left(x, y; Q^2, Q_0^2 \right) &=&
\frac{1}{2} \int_0^1 dz\; G(x,y,z|\nu(A))
          U_{\rm DGLAP}\left(z, z'=1, \eta=0; Q^2, Q_0^2 \right),
\end{eqnarray}
where $G(x, y; z| \nu)$  is the kernel defined in Eq.\ (\ref{Def-IntKerG}).
Beyond the LO approximation, one can perturbatively expand the solution with
the help of this evolution operator. From the representation of this
operator in terms of Gegenbauer polynomials it is obvious that the inverse
operator is $U^{-1}_{\rm ERBL}\left(x, y; Q^2, Q_0^2 \right)= U_{\rm
ERBL}\left(x, y; Q_0^2, Q^2\right)$.

Let us add that in the forward case, i.e. $\langle P_2| \to \langle P_1|$,
Eqs.\ (\ref{def-SPD-tra}) define the chiral odd quark density and the
tensor gluon density times the momentum fraction $z$, thus, we have the
limit of the splitting
functions:
\begin{eqnarray}
\label{Def-LIM}
\left\{ {{^{QQ}\!P^T} \atop {^{GG}\!P^T}} \right\} (z)
=
{\rm LIM}
\left\{ {{^{QQ}V^T} \atop {^{GG}V^T}} \right\} (x, y)
\equiv \lim_{\eta\to 0} \frac{1}{|\eta|}
\left\{ {{^{QQ}V^T} \atop \frac{1}{z}{^{GG}V^T}} \right\}
\left( \frac{z}{\eta}, \frac{1}{\eta} \right) ,
\end{eqnarray}
which will be of relevance to the analysis in the following section.

%%%%%%%%%%%%%%%%%%%%%%%%%%%%%%%%%%%%%%%%%%%%%%%%%%%%%%%%%%%%%%%%%%%%%
\section{Construction of two-loop kernels.}
\label{sec-rec}
%%%%%%%%%%%%%%%%%%%%%%%%%%%%%%%%%%%%%%%%%%%%%%%%%%%%%%%%%%%%%%%%%%%%%

Since the extension of the ER-BL kernel to the whole region is unique
\cite{MueRobGeyDitHor94DitGeyMueRobHor88}, it is sufficient to construct
the evolution kernels,
\begin{equation}
{^{AA}V^T} (x, y)
= \frac{\alpha_s}{2\pi} {^{AA}V^{T(0)}} (x, y)
+ \left( \frac{\alpha_s}{2\pi} \right)^2
{^{AA}V^{T(1)}} (x, y) + {\cal O} \left( \alpha_s^3 \right) ,
\end{equation}
in the ER-BL-region. The one loop kernels have been obtained by a direct
calculation \cite{BukFroKurLip85,BelMue97a,HooJi98,BelMue00}.
Alternatively, they can be deduced by means of conformal covariance
(\ref{eigenvec}) from the forward kernels via the integral transformation
(\ref{Def-IntKerG}):
\begin{equation}
\label{Def-ERBL-LO-tr}
\left\{ { {^{QQ}V^{T(0)}} \atop {^{GG}V^{T(0)}} } \right\}
= \left\{ { C_F \atop C_A } \right\}
\left[ \theta(y - x)
\left\{ { {^{QQ}\!f^T} \atop {^{GG}\!f^T} } \right\}
+ \Bigg\{ {x \to \bar x \atop  y \to \bar y } \Bigg\} \right]_+
- \frac{1}{2}
\left\{ { {^{QQ}\gamma^{T(0)}_0} \atop {^{GG}\gamma^{T(0)}_1} } \right\}
\delta(x - y) ,
\end{equation}
where
\begin{equation}
\label{B-kernel-f}
{^{QQ}\!f^T} \equiv {^{QQ}\!f^b} = \frac{x}{y} \frac{1}{y - x} ,
\quad
{^{QQ}\gamma^{T(0)}_0} = C_F ,
\quad
{^{GG}\!f^T} \equiv {^{GG}\!f^b} =  \frac{x^2}{y^2} \frac{1}{y - x},
\quad
{^{GG}\gamma^{T(0)}_1} = 6 C_A + \beta_0 .
\end{equation}
Here the +-prescription is defined for both channels in the same way,
namely as $[V(x,y)]_+ = V(x,y) - \delta(x-y) \int_{-1}^1 dz V(z,y)$.
As explained in great detail in \cite{Mue94BelMue98aBelMue98c}, in NLO
conformal covariance does not hold true anymore for the minimal subtraction
(MS) scheme in the dimensional regularized theory. Besides a symmetry
breaking term proportional to the QCD $\beta$-function, there appears
an additional term which is induced by the leading order anomaly in the
special conformal transformations of conformal operators. Note, that the
latter anomalies are renormalization scheme dependent even in one-loop
approximation. Next, the results for the local conformal operators in the
MS scheme can be transformed to the kernels, which have the following
structure in NLO:
\begin{eqnarray}
\label{Def-NLO-Ker}
{^{AA}V^{T(1)}}
= - {^{AA} \dot V^T}
\otimes \left( {^{AA} V^{T(0)}} + \frac{\beta_0}{2} \1 \right)
- \left[{^{AA}\!g^T} \OO_{\mbox{'}} {^{AA}V^{T(0)}} \right]
+ {^{AA}{\cal D}^T} ,
\end{eqnarray}
where the commutator stands for $[ A \, {\displaystyle \OO_{\mbox{'}} } \,
B ] (x, y) = \int_0^1 dz\, \left\{ A (x, z) B (z, y) - B (x, z) A (z, y)
\right\}$. The off-diagonal part w.r.t.\ Gegenbauer polynomials is
contained in the first two convolutions on the r.h.s.\ of this equation.
The third term, i.e.\ ${^{AA}{\cal D}^T}$, is diagonal w.r.t.\ Gegenbauer
polynomials with index $\nu(A) = \left\{ \ft32, \ft52 \right\}$ for $A =
\{ Q, G \}$. The dotted kernel is derived from the LO ones via a logarithmic
modification,
\begin{eqnarray}
{^{QQ}\dot V^{T(0)}}
&=& C_F \theta(y - x) \frac{x}{y} \frac{1}{y - x} \ln\frac{x}{y}
+ \left\{x \to \bar x \atop y \to \bar y \right\},
\\
{^{GG}\dot V^{T(0)}}
&=& C_A \theta(y - x) \frac{x^2}{y^2} \frac{1}{y - x} \ln\frac{x}{y}
+ \left\{x \to \bar x \atop y \to \bar y \right\} .
\end{eqnarray}
The $g$-kernels, meanwhile, contain new information and are defined by
\begin{eqnarray}
{^{QQ}\!g^T}
\!\!\!&=&\!\!\! - C_F \left[ \theta(y - x)
\frac{\ln\left(1 - \frac{x}{y}\right)}{y - x}
+ \left\{x \to \bar x \atop y \to \bar y \right\} \right]_+,
\\
{^{GG}\!g^T}
\!\!\!&=&\!\!\! - C_A \left[\theta(y - x)
\left( \frac{\ln\left(1 - \frac{x}{y}\right)}{y - x} - 2 \frac{x}{y} \right)
+ \left\{x \to \bar x \atop y \to \bar y \right\} \right]_+ .
\end{eqnarray}
In contrast to the $QQ$-kernel, which is unique in all sectors
\cite{Mue94BelMue98aBelMue98c}, the $GG$-kernel differs from those in the
chiral even case by the term $2 C_A \ft{x}{y}$ \cite{BelMue00}.

Since it is difficult to project onto the off-diagonal part and apply
the integral transformation to the NLO DGLAP kernel \cite{Vog98}, we use
another strategy for the construction of the remaining diagonal term,
${\cal D}$. This term is decomposed into a contribution related to the
crossed-ladder diagram containing dilogarithms, $G$, and a remainder $D$
which is, for the present helicity-flip sector, given as a linear
combination of one-loop kernels,
\begin{eqnarray}
\label{Def-RemDia}
\left\{ { {^{QQ}{\cal D}^T} \atop {^{GG}{\cal D}^T} } \right\} (x, y)
= - \frac{1}{2}
\left\{ { C_F \left(C_F - \frac{C_A}{2} \right)
\left[ {^{QQ}G^T} (x, y) \right]_+
\atop
C_A^2 \left[ {^{GG}G^T} (x, y) \right]_+ } \right\} +
\left\{ { {^{QQ}\!D^T} \atop {^{GG}\!{D}^T} } \right\} (x, y) .
\end{eqnarray}
In the following the $G^T$ kernels are defined as
\begin{equation}
\label{Def-G}
{^{AA}\! G^T}(x,y)
= \theta (y - x) \left( {^{AA}h^T} + \Delta {^{AA}h^T} \right) (x, y)
+ \theta (y - \bar x) \left( {^{AA}\bar h^T}
+ \Delta {^{AA}\bar h^T} \right) (x, y) ,
\end{equation}
where
\begin{eqnarray}
\label{Def-Func-h}
{^{AA}h^T}
\!\!\!&=&\!\!\!
2 \, {^{AA}\bar f^T} \ln \bar x \ln y
- 2 \, {^{AA}\!f^T}
\left[ {\rm Li}_2 (x) + {\rm Li}_2 (\bar y) \right] ,
\\
{^{AA}\bar h^T}
\!\!\!&=&\!\!\!
\left( {^{AA}\!f^T} - {^{AA}\!\bar f^T} \right)
\left[ 2 {\rm Li}_2 \left( 1 - \frac{x}{y} \right) + \ln^2 y \right]
+ 2 \, {^{AA}\!f^T} \left[ {\rm Li}_2 ( \bar y ) - \ln x \ln y \right]
+ 2 \, {^{AA}\!\bar f^T} {\rm Li}_2 ( \bar x ) .
\nonumber
\end{eqnarray}
Here the $QQ$ kernel follows from the explicit flavor non-singlet
two-loop result in the chiral even sector \cite{Sar84} by replacing
${^{QQ}\!f^{\rm NS}} \to {^{QQ}\!f^T}$, with the addenda $\Delta
{^{QQ}h^T}$ and $\Delta {^{QQ}\bar h^T}$ being zero. Note that we
rewrote the original result in such a way that both $\theta$-contributions
are symmetric with respect to the weight function $x \bar x$, i.e.\ $y
\bar y {^{QQ}h^T} (x, y) = x \bar x {^{QQ}h^T} ( \bar y, \bar x )$ and
$y \bar y {^{QQ}\bar h^T} (x, y) = x \bar x {^{QQ}\bar h^T} (y, x)$. Now
it can be verified that both parts are separately diagonal w.r.t.\
Gegenbauer polynomials.

Since the two-loop crossed-ladder diagram contains only a simple pole in
the dimensional regularization parameter, the ${\cal N} = 1$ supersymmetric
constraints on the level of local operators lead to the equality of chiral odd
quark and tensor gluon anomalous dimensions ${^{QQ}\gamma^{T +}_j} =
{^{GG}\gamma^T_j}$ \cite{BukFroKurLip85,BelMue99proc}. From this constraint 
one can derive a relation between the momentum fraction kernels which reads 
[for details see \cite{BelFreMue00BelMueFre99}]:
\begin{eqnarray}
\frac{\partial}{\partial y} {^{QQ}G^T} (x, y)
+
\frac{\partial}{\partial x} {^{GG}G^T} (x, y) = 0 .
\end{eqnarray}
Taking the generic form (\ref{Def-G}) for ${^{AA}G^T} (x, y)$, this
differential equation and the symmetry requirements imply the following
form of the gluonic addenda (the remaining degree of freedom is fixed
by the diagonality of the lowest moments):
\begin{equation}
\label{Def-Add-Tr}
\Delta{^{GG}\!h^T} (x, y) = - \frac{2 x}{y^2 \bar y}
- \frac{2 \bar x}{y^2 \bar y} \ln \bar x - \frac{2 x}{y \bar y^2} \ln y ,
\qquad
\Delta{^{GG}\bar h^T} (x, y) = \Delta{^{GG}h^T} (\bar x, y) .
\end{equation}
The missing diagonal ${^{AA}\!{D}^T}$ terms are easily obtained in the
DGLAP representation by taking the forward limit and comparison with
the two-loop results \cite{Vog98}, namely,
\begin{equation}
\label{get-D}
{^{AA}\!D^T} (z) = {^{AA}\!\dot P^T} (z)
- {\rm LIM}
\left\{
- {^{AA}\dot V^T} \OO^\re
\left( {^{AA}V^{(0)T}} + \frac{\beta_0}{2} \1 \right)
- \left[ {^{AA}\!g^{T}} \OO^\re_{,} {^{AA}V^{(0)T}} \right]_-
-\frac{1}{2} G\mbox{-terms}
\right\} .
\nonumber
\end{equation}
Since the result is expressed in terms of LO splitting functions, we can
easily restore the ER-BL representation from them\footnote{Note that the
last term in the $QQ$ channel was missed in the second article of
Ref.\ \cite{BelFreMue00BelMueFre99}.}:
\begin{eqnarray}
\label{Cal-Dia-Rem}
{^{QQ}\!D^{T}} (x, y) = \!\!\!&-&\!\!\!
C_F \left( \frac{2}{3} C_F + \frac{5}{6} \beta_0 \right)
{^{QQ}\!v^b} (x, y) \nonumber\\
&-&\!\!\!\ C_F \left( C_F - \frac{C_A}{2} \right)
\left\{ - {^{QQ}\!v^a} (x, y) - {^{QQ}\!v^a}(\bar x, y)
+ \frac{4}{3} {^{QQ}\!v^b} (x, y) \right\} ,
\\
{^{GG}\!D^{T}} (x, y) = \!\!\!&-&\!\!\! C_F T_F N_f
\left\{ {^{GG}\!v^a} + \frac{2}{3} {^{GG}\!v^c} \right] (x, y)
+ \beta_0 C_A \left\{ \frac{3}{8} {^{GG}\!v^a} - \frac{5}{6} {^{GG}\!v^b}
+ \frac{1}{4} {^{GG}\!v^c} \right\} (x, y)
\nonumber\\
&+&\!\!\! C_A^2 \left\{ \frac{13}{8} {^{GG}\!v^a}
- \frac{11}{6} {^{GG}\!v^b} + \frac{13}{12} {^{GG}\!v^c} \right\} (x, y) ,
\nonumber
\end{eqnarray}
where the $b$-kernel has been defined above in Eq.\ (\ref{B-kernel-f})
and the $a$- and $c$-kernels are given by
\begin{eqnarray}
{^{QQ}\!v^a} = \frac{x}{y} ,
\qquad
{^{GG}\!v^a} = \frac{x^2}{y^2} ,
\qquad
{^{GG}\!v^c} = \frac{x^2}{y^2}\left(2 \bar x y + y - x \right) .
\end{eqnarray}
Contributions concentrated in $x = y$ will be restored below.

It remains to perform the exclusive convolution. For two non-regularized
kernels given by their functions $f_i (x, y)$ with $i = \{ 1, 2 \}$ we
have
\begin{equation}
\left( v_1 \otimes v_2 \right) (x, y)
= \theta(y - x) \left( f_1 \otimes f_2 \right) (x, y)
+ \left\{ { x \to \bar x \atop y \to \bar y} \right\} ,
\end{equation}
where
\begin{equation}
\left( f_1 \otimes f_2 \right) (x, y)
= \int_x^y dz \, f_1 (x, z) f_2(z, y)
+ \int_y^1 dz \, f_1 (x, z) f_2 (\bar z, \bar y)
+ \int_0^x dz \, f_1 (\bar x, \bar z) f_2 (z, y) .
\end{equation}
The convolution of two regularized kernels is slightly more involved.
Here we define:
\begin{eqnarray}
\left( \left[ v_1 \right]_+ \otimes \left[ v_2 \right]_+ \right) (x, y)
=
\left[ \theta(y - x) \left( f_1 \otimes f_2 \right) (x, y)
+ \left\{ { x \to \bar x \atop y \to \bar y } \right\} \right]_+ ,
\end{eqnarray}
where the convolution on the r.h.s.\ can be decomposed into well-defined
integrals
\begin{eqnarray}
\left( f_1 \otimes f_2 \right) (x, y)
\!\!\!&=&\!\!\! \int^y_x dz
\left\{
\left[ f_1 (x, z) - f_1 (x, y) \right]
\left[ f_2 (z, y) - f_2 (x, y) \right]
+
\left[ f_1 (x, z) - f_1 (\bar z, \bar x) \right] f_2 (x, y)
\right\}
\nonumber\\
&+&\!\!\! \int^1_y dz
\left\{
\left[ f_1 (x, z) - f_1 (x, y) \right] f_2 (\bar z, \bar y)
-
\left[ f_1 (\bar z, \bar x) - f_1 (x, y) \right] f_2(x, y)
\right\}
\nonumber\\
&+&\!\!\! \int^x_0 dz
\left\{
\left[ f_1 (\bar x, \bar z) - f_1 (x, y) \right]
\left[ f_2 (z, y) - f_2 (x, y) \right]
+
\left[ f_1 (\bar x, \bar z) - f_1 (z, x) \right] f_2 (x, y)
\right\}
\nonumber\\
&-&\!\!\! f_1 (x,y) f_2 (x, y).
\end{eqnarray}
Performing the exclusive convolution in Eq.\ (\ref{Def-NLO-Ker})
and adding the diagonal terms given in Eqs.\ (\ref{Def-RemDia})--
(\ref{Def-Func-h}), (\ref{Def-Add-Tr}),  and
(\ref{Cal-Dia-Rem}), the $QQ$ and $GG$ transversity kernels in NLO
are found to be:
\begin{eqnarray}
{^{QQ}V^{(1)T\pm}}
\!\!\!&=&\!\!\! \left[
\theta (y - x) {^{QQ}\!{\cal F}^{T\pm}} (x, y) 
+ C_F^2 \, \frac{\ln \bar x \ln x}{y}
+ \left\{ { x \to \bar x \atop y \to \bar y } \right\}
\right]_+
- \frac{1}{2} {^{QQ}\gamma_0^{(1)T\pm}} \delta(x - y) , \\
{^{GG}V^{(1)T}}
\!\!\!&=&\!\!\! \left[
\theta (y - x) {^{GG}\!{\cal F}^T} (x, y) 
+ C_A^2 \left\{
\frac{x (\bar y - y)}{ y \bar y} \ln x
+ \frac{x + y}{y^2} \ln x \ln \bar x
\right\}
+ \left\{ { x \to \bar x \atop y \to \bar y } \right\}
\right]_+ \nonumber\\
&&\qquad\qquad\qquad\qquad\qquad\qquad\qquad\qquad\qquad\qquad\qquad
- \frac{1}{2} {^{GG}\gamma_1^{(1)T}} \delta(x - y) ,
\end{eqnarray}
where
\begin{eqnarray}
\label{Def-ERBL-NLO-trQ}
&&{^{QQ}\!{\cal F}^{T\pm}} (x, y) \nonumber\\
&&\qquad= C_F^2 \Bigg\{
2 \left( \frac{2}{3} - \zeta(2) \right) \qqv
- \qqv \left( \frac{3}{2} - \ln \frac{x}{y} \right) \ln \frac{x}{y}
- ( \qqv - \qqbv ) \ln \frac{x}{y} \ln \left( 1 - \frac{x}{y} \right)
\Bigg\} \nonumber\\
&&\qquad- C_F \left( C_F - \frac{C_A}{2} \right)
\left\{ -( 1 \mp 1 ) \qqva
+ \frac{4}{3} \qqv + \HQQ^T (x, y) \mp \bHQQ^T (\bar x, y) \right\}
\nonumber\\
&&\qquad- \frac{1}{2} C_F \beta_0
\left( \frac{5}{3} + \ln \frac{x}{y} \right) \qqv , \\
\label{Def-ERBL-NLO-trG}
&&{^{GG}\!{\cal F}^T} (x, y) \nonumber\\
&&\qquad=
C_A^2 \Bigg\{
2 \left( \frac{1}{3} - \zeta(2) \right) \ggvv
+ \ggvv \ln^2 \frac{x}{y}
- \left( \ggvv - \ggbvv \right)
\ln \frac{x}{y} \ln \left( 1 - \frac{x}{y} \right) \nonumber\\
&&\qquad
+ \frac{13}{4} \left( \frac{1}{3} \ggvc + \frac{1}{2} \ggva \right)
- \frac{1}{2} \HGGe (x, y) - \frac{1}{2} \bHGGe (\bar x, y)
- \Delta \HGGe (x, y) \Bigg\} \nonumber\\
&&\qquad+ \frac{1}{2} C_A \beta_0
\Bigg\{
\frac{3}{4} \ggva - \frac{5}{3} \ggvv + \frac{1}{2} \ggvc
\Bigg\}
- C_F N_f T_F \Bigg\{ \ggva + \frac{2}{3} \ggvc \Bigg\} .
\end{eqnarray}
The contributions concentrated on the diagonal $x = y$ are fixed by the
lowest conformal moment. We have in the $QQ$ and $GG$ channels:
\begin{eqnarray}
{^{QQ}\!\gamma^{(1)T\pm}_{0}} &=&
\frac{19}{12} C_F^2 - \frac{13}{12} C_F \beta_0 -
C_F\left(C_F-\frac{C_A}{2}\right)
\left(\frac{19}{3} +(1\pm 1)
\left[7 - 8\zeta(2) + 4\zeta(3)\right]\right) ,
\nonumber\\
{^{GG}\!\gamma^{(1)T}_{1}} &=&
  \frac{19}{4} C_A^2 - \frac{9}{4} C_A N_f + \frac{3}{2} C_F N_f.
\end{eqnarray}

Finally, we derive the NLO skewed DGLAP kernels in Radyushkin's notation
with fraction $z = \frac{x + \eta}{1 + \eta}$ (here $x$ denotes the
momentum fraction in the SPDs) and skewedness $\zeta = \frac{2\eta}{1 +
\eta}$. These kernels govern the evolution of SPDs in the kinematic range
of $\zeta < z < 1$ and $- 1 + \zeta < z < 0$. We choose this set of
variables such that we have the closest resemblance to the usual DGLAP
kinematics. In fact for DVCS and vector meson production $\zeta =
x_{\rm Bj}$ up to terms $\sim \frac{\Delta^2}{Q^2}$! The derivation
procedure follows that of the forward limit reduction of the extended
ER-BL kernels. After the analytical continuation of the ER-BL kernels
to the whole $\{x, y \}$ plane by the replacement of the $\theta$-function
structure, we derive the kernels in the following way which is suggested
by the non-zero support of the $\theta$-functions after the replacement
$x \rightarrow \ft{z}{\zeta}$, $\bar x \rightarrow 1 - \ft{z}{\zeta}$,
$y \rightarrow \ft1{\zeta}$ and $\bar y \rightarrow 1 - \ft1{\zeta}$ for
the above mentioned kinematical regime:
\begin{equation}
P^T (z, \zeta)
= \frac{1}{\zeta}
\left\{
{^{AA}V^T} \left( \frac{z}{\zeta}, \frac{1}{\zeta} \right)
-
{^{AA}V^T} \left( 1 - \frac{z}{\zeta}, 1 - \frac{1}{\zeta} \right)
\right\} .
\end{equation}
Note again that we omitted the additional factor of $z^{-1}$ in the
$GG$ kernel so as to take into account that in the forward
limit, the skewed gluon distribution turns into $z G (z, Q^2)$ rather
than $G (z, Q^2)$.

The results for the chiral odd DGLAP kernels in LO, following our above
prescription, are:
\begin{equation}
{^{QQ}p^T}
= \frac{1 + z}{\bz} - \frac{1}{\be},
\qquad
{^{GG}\!p^T} = \frac{(1 + z) z}{\bz} + \frac{1 - \zeta z}{\be^2}
- \frac{1 + \zeta + z}{\be} ,
\end{equation}
and in NLO:
\begin{eqnarray}
{^{QQ}P^{T(1)\pm}} (z, \zeta)
\!\!\!&=&\!\!\! C_F^2 \Bigg\{\frac{1}{2} \left( \pqqt - \frac{\zeta}{\be} \right)
\loembze \left[ \loembze - \frac{3}{2} \right]
+ \frac{z}{\bz} \loz \left( \loz - \frac{3}{2} \right) \nonumber\\
&&- \pqqt \left[ \loembze \lobzobe + \lobz \loz \right]
+
\left( \frac{4}{3} - 2 \zeta(2) \right) \pqqt
\Bigg\} \nonumber\\
&-&\!\!\! \frac{1}{2} C_F \beta_0
\Bigg\{ \frac{5}{3} \pqqt + \frac{z}{\bz} \loz
+ \left( \frac{1}{\bz} - \frac{1}{\be} \right)\loembze \Bigg\}
\nonumber\\
&-&\!\!\! C_F \left( C_F - \frac{C_A}{2} \right)
\Bigg\{
- (1 \mp 1) \pqqa + \frac{4}{3} \pqqt
+ {^{QQ}h^T} \mp {^{QQ}\bar h^T} \Bigg\} , \\
{^{GG}\!P}^T (z, \zeta) \!\!\!&=&\!\!\!
C^2_A \Bigg\{
\frac{1}{2} \lozembze
\left( \pggpt - \frac{ \bz + \zeta(1 + z)}{\be} z + \frac{\bz}{\be^2} \right)
+ \frac{13}{4} \left( \frac{1}{3} \pggc + \frac{1}{2} \pgga \right)
\nonumber\\
&&- 2 \frac{\zeta}{\be}
\Bigg( \zeta \bzoe + \frac{z}{\be} \loz - \be \embze \loembze \Bigg)
+ \frac{z^2}{\bz}\lozz
+ \left( \frac{2}{3} - 2 \zeta(2) \right) \pggpt \nonumber\\
&&- \pggpt \left( \loembze \lobzobe + \lobz \loz \right)
- \frac{1}{2} {^{GG}h^T}  - \frac{1}{2} {^{GG}{\bar h}^T} \Bigg\}
\nonumber\\
&+&\!\!\! \frac{1}{2} C_A \beta_0
\Bigg\{ \frac{3}{4} \pgga - \frac{5}{3} \pggpt + \frac{1}{2} \pggc \Bigg\}
- C_F N_f T_F \Bigg\{ \pgga + \frac{2}{3} \pggc \Bigg\} ,
\end{eqnarray}
where one has to regularize the {\it whole} kernel with the $+$-prescription
and add the end-point contribution $-\frac{1}{2} {^{QQ}\!\gamma_0^{(i)
T\pm}} \delta(1 - z)$ and $-\frac{1}{2} {^{GG}\!\gamma_i^{(i)T}}
\delta(1 - z)$ specified earlier. The ${^{AA}h^T}$ and ${^{AA}\bar h^T}$
functions are given by the equations
\begin{eqnarray}
{^{AA}h^T} \!\!\!&=&\!\!\! 2 \, {^{AA}p^T} (z, \zeta)
\left[ \ln \zeta \loembze
+ {\rm Li}_2 \left( 1 - \frac{1}{\zeta} \right)
- {\rm Li}_2 \left( 1 - \zoe \right) - \zeta (2) \right]
\nonumber\\
&+&\!\!\! 2 \, {^{AA}k_1^T} (z, \zeta)
\left[
\loembze \left( \ln \left( \zoe \right) - \ln \zeta \right)
- 2 \left( {\rm Li}_2 \left( 1 - \frac{1}{\zeta} \right)
- {\rm Li}_2 \left( 1 - \zoe \right) \right)
\right]
\nonumber\\
{^{AA}\bar h^T} \!\!\!&=&\!\!\! 2 \, {^{AA}p^T} (\zeta - z, \zeta)
\Bigg[ \ln \zeta \loembze
+ \frac{1}{2} \loe \left( \ln \left( \zoe \right) + \loz
- \ln \left( \frac{\be}{\zeta} \right) \right)
+ {\rm Li}_2 \left( 1 - \frac{1}{\zeta} \right)
\nonumber\\
&&- {\rm Li}_2 \left( 1 - \zoe \right) - \zeta(2)
- \ln(\be + z) \left( \loembze + \loz \right) - {\rm Li}_2 (\zeta - z)
- {\rm Li}_2 \left( - \frac{z}{\be} \right) \Bigg]
\nonumber\\
&+&\!\!\! 2 \, {^{AA}k_2^T} (z, \zeta)
\left[ \loembze \left( \ln \left( \zoe \right) - \ln \zeta \right)
- 2 \left( {\rm Li}_2 \left( 1 - \frac{1}{\zeta} \right)
- {\rm Li}_2 \left( 1 - \zoe \right) \right) \right] .
\end{eqnarray}
The remaining skewed kernels appearing in the above formulae read
\begin{equation}
{^{QQ}p^a} = \frac{\bz}{\be} \, ,
\qquad
{^{GG}\!p^a} = 2 \frac{z \bz}{\be^2} - \frac{\zeta}{\be^2} (1 - z^2) \, ,
\qquad
{^{GG}p}^c = \frac{\bz^3}{\be^2} \, ,
\end{equation}
and the ${^{AA}k^T_i}$ were found to be:
\begin{equation}
{^{QQ}k^T_1} = \frac{z}{\bz} \, ,
\qquad
{^{QQ}k^T_2} = -\frac{z}{\be(\be + z)} \, ,
\qquad
{^{GG}\!k^T_1} = \frac{z^2}{\bz} \, ,
\qquad
{^{GG}\!k^T_2} = - \frac{z^2}{\be^2(\be + z)} .
\end{equation}
Note, that we have included the addenda of the skewed $G$-functions in
the single logs and the rational functions of the respective colour
structures in the kernels so as to simplify the overall presentation of
the results.

%%%%%%%%%%%%%%%%%%%%%%%%%%%%%%%%%%%%%%%%%%%%%%%%%%%%%%%%%%%%%%%%%%%%%
\section{Conclusion.}
\label{sec-con}
%%%%%%%%%%%%%%%%%%%%%%%%%%%%%%%%%%%%%%%%%%%%%%%%%%%%%%%%%%%%%%%%%%%%%

In conclusion, we have presented a calculation of the two-loop exclusive
evolution kernels for skewed chiral odd quark and tensor gluon distributions,
thus allowing for NLO analysis of the cross sections where these functions
contribute. Our modus operandi is based on the formalism developed
previously in Refs.\ \cite{BelFreMue00BelMueFre99} for vector and axial
sectors. Due to chirality conservation only the helicity flip gluon SPD shows
up in the DVCS cross section. A peculiar azimuthal angle dependence of
asymmetries on the nucleon target \cite{BelMue00} will give an opportunity
to pin down the magnitude of the gluon SPD provided high accuracy data
will be available. This will shed some light on the unknown tensor gluon
content of the nucleon or higher spin hadrons.

\vspace{0.5cm}

A.B. would like to thank A. Sch\"afer for the hospitality extended
to him at the Institut f\"ur Theoretische Physik, Universit\"at
Regensburg. This work was supported by the Alexander von Humboldt Foundation,
in part by the National Science Foundation, under grant PHY9722101 and
Graduiertenkolleg Erlangen-Regensburg (A.B.), by the DFG and the BMBF (D.M.),
and by the E.U. contract FMRX-CT98-0194 (A.F.).

\end{document}